# Unimib Assistant: designing a student-friendly RAG-based chatbot for all their needs

Chiara Antico[1,*], Stefano Giordano[1,*], Cansu Koyuturk[1] and Dimitri Ognibene[1]

[1] *Dept. Psychology, Università degli Studi di Milano Bicocca, Milan, Italy*

**Abstract**
Natural language processing skills of Large Language Models (LLMs) are unprecedented, having wide diffusion and application in different tasks. This pilot study focuses on specializing ChatGPT behavior through a Retrieval-Augmented Generation (RAG) system using the OpenAI "custom GPTs" feature. The purpose of our chatbot, called "Unimib Assistant", is to provide information and solutions to the specific needs of University of Milano-Bicocca (Unimib) students through a question-answering approach. We provided the system with a prompt highlighting its specific purpose and behavior, as well as university-related documents and links obtained from an initial need finding phase, interviewing six students. After a preliminary customization phase, a qualitative usability test was conducted with six other students to identify the strengths and weaknesses of the chatbot, with the goal of improving it in a subsequent redesign phase. While the chatbot was appreciated for its user-friendly experience, perceived general reliability, well-structured responses, and conversational tone, several significant technical and functional limitations emerged. In particular, the satisfaction and overall experience of the users was impaired by the system's inability to always provide fully accurate information. Moreover, it would often neglect to report relevant information even if present in the materials uploaded and prompt given. Furthermore, it sometimes generated unclickable links, undermining its trustworthiness, since providing the source of information was an important aspect for our users. Further in-depth studies and feedback from other users as well as implementation iterations are planned to refine our Unimib Assistant.

## 1. Introduction

The advent of Generative Artificial Intelligence (GenAI) and Large Language Models (LLMs) such as ChatGPT by OpenAI, which can learn patterns and structures from its training data to generate novel content, has allowed conversational agents to respond in a more flexible, dynamic and context-aware manner than traditional rule-based chatbots [1]. Since its launch in November 2022, ChatGPT has gained over 180 million registered users and has received 600 million monthly visits as of May 2024 [2]. It is capable of handling diverse queries across multiple topics, engaging in casual conversations using natural language, generating coherent text, accounting for human feedback, and adapting its responses to enhance user interaction [3]. Despite its capabilities, LLMs are not without flaws. They can produce responses that appear correct but are in fact nonsensical or inaccurate, a phenomenon referred to as "hallucinations" [4]. This can lead to misinformation, as users may also struggle to distinguish what constitutes correct information, particularly during information search. Outdated knowledge is another significant limit of these systems, undermining users' trust in the reliability of their outputs, especially in question-answering tasks [5]. A possible method to address these issues is employing a Retrieval-Augmented Generation (RAG) system that combines information retrieval with generative models to enhance content generation by providing relevant, up-to-date information from external data sources to be processed by the LLMs together with the user request [6]. This helps to improve the accuracy and contextual relevance of generated content, as well as the ability to incorporate extensive knowledge and update information more easily to address challenges such as outdated, irrelevant, or misleading knowledge [7]. In addition, RAG systems are particularly valuable for tailoring responses to specific contexts where the required information might not be available to the LLM or for prioritizing certain information over others, ensuring greater relevance to the user's needs. Despite the potential of these systems, they are still constrained by the inherent limitations of LLMs and the complexities involved in the information retrieval process [8].

## 1.1. Related works

An effective prompt design, that is, a series of instructions and guidelines for the chatbot to follow, is not only an essential technical consideration but also a fundamental aspect both to create chatbots that can engage users in meaningful conversations and to affect the flow and quality of conversations [9]. Particularly novice users can create prompts that are either too broad or overly specific which limits the effectiveness of the chatbot's behavior and can lead to challenges in generating desired outputs [10]. On the other hand, AI agents can exhibit unnatural and highly structured response formats, limited coverage of topics, and unreliable behaviors that could disrupt the flow of the interactions [11]. For example, a study by Gravel and colleagues found that 69% of the references provided by ChatGPT in response to medical questions were fabricated [12]. Theophilou and colleagues [16] suggest that educational activities involving hands on experience with ChatGPT could improve users' prompting skills and awareness of LLMs' capabilities and limits.

There are studies that have explored knowledge-augmented LLM frameworks to improve chatbot performance and the User Experience (UX). Baek and colleagues showed that by incorporating users' historical search interactions and browsing activities, a knowledge-augmented chatbot significantly enhanced the relevance and usefulness of query recommendations, especially for users with extensive search histories, thus enhancing their experience with the system [13]. Thway and colleagues [14] investigated a RAG-based chatbot designed to deliver personalized, on-demand course content. It was able to provide up-to-date and domain-specific information, which was found to increase student engagement with the course material and improve academic performance, with an overall beneficial effect on the UX. Similarly, Jacobs and Jaschke found that the RAG-enabled chatbot providing feedback on programming tasks could provide relevant lecture content to help students better understand concepts, although there was a tradeoff between the depth and speed of feedback [15]. The integration of RAG systems has also been shown to enhance the relevance of answer for the user queries, which refers to the action of seeking information through question asking, though this has some limitations. Mansurova and colleagues [5] designed a chatbot to enhance question-answering capabilities by leveraging the QA-RAG framework. They found that while the chatbot achieved rather high accuracy in various RAG capabilities, it still struggled with integrating external data that may contradict its pre-existing knowledge, with an inevitable impact on the UX also to be considered. Finally, LLM based chatbots, even when augmented with information retrieval mechanisms, have still significant limitations in asking refining questions to provide more accurate responses [21, 22].

## 1.2. Review of student-friendly technological support

Before starting the customization of the chatbot, which aims to support students at the University of Milano Bicocca (Unimib), we reviewed on July-August 2024 the already existing technological products that Bicocca, and universities in general, use to support their students.

Unimib offers different applications to allow students to easily stay updated about important information, such as the Unimib Course app, which provides information mainly on lessons and exams, classrooms and timetables, and the BicoccApp, which manages one's academic career. Also, social media pages are frequently used to provide students with important information and knowledge concerning any university topic: from administrative and bureaucratic deadlines, Erasmus calls, scholarships, university news and ongoing projects, to students' rights, activist demonstrations, special events, student representatives' elections, to name some.

Usually universities, including Unimib, use several other technological methods to support their students in looking for information, e.g., official websites[2] where students can either proactively search for information through the magnifier icon or explore the portal by looking at each specific section and subsections present; intranet platform, which is the secure and immediate access tool to the university's internal resources and services. The intranet[3] is a virtual workspace dedicated to the university's teaching, research and technical-administrative staff, designed primarily to make the organization's internal communication process more immediate and effective, supporting the activities of all departments; Moodle environment is also a good source of information, such as the e-learning platform[4], which is a didactic tool, with full web access and use, that supports traditional classroom teaching and allows the lecturer to publish and make lecture material accessible to students, to transmit communications, to publish course and lecture information, to administer assignments/exams, tests and more.

When designing and integrating LLM-based solutions into complex and diverse information systems, it is essential to consider the varying update frequencies and structural organization of sources. Infrequently updated, high-importance documents may benefit from fine-tuning the LLM, while frequently updated sources are better suited for a RAG system to ensure access to current data without constant retraining. Additionally, highly structured content may require specifically designed retrieval systems or tailored interfaces to optimize the interaction between the LLM and the data. Finally, practical testing of how the RAG-LLM system processes and extracts information from these sources to address user requests is vital, as its performance can be difficult to predict in advance.

## 2. Description of the system

The main difference between all the methods cited above and our system is that navigating and searching between the different university sites and platforms is not a trivial problem for students. They can sometimes become confused and disoriented if they are not familiar with these platforms and, sometimes, they may retrieve redundant or outdated information.

Therefore, our aim is to develop an efficient LLM-based chatbot, called *Unimib Assistant*, designed to provide students with quick, relevant, easy to understand, and reliable information in response to their questions on university topics. We do not intend to replace the existing products the university puts in the service of its students (i.e., the official platforms and existing applications), but rather to encompass all this knowledge into our chatbot, whose integration of consistent human-like responses and the retrieval of information from the sources uploaded will meet and satisfy the needs of our users. We further believe that our system can help students better familiarize themselves with the university, its features, services, and opportunities by allowing them to ask the bot for any type of information. In addition, with our virtual assistant, students may also learn how to navigate the various university platforms and find information within them, providing direct links to the webpage of interest.

---

[2] Home. Università degli Studi di Milano-Bicocca. (2024, April 10). https://www.unimib.it/
[3] La Intranet di Ateneo si rinnova. Università di Torino (2021, June 29). https://www.unito.it/avvisi/la-intranet-di-ateneo-si-rinnova
[4] Learning edidattica online. e. (2024, May 28). https://elearning.unimib.it/

ChatGPT-4 enables the creation of a personalized virtual assistant, a customized GPT, by combining the generative power of LLMs with the retrieval of specific information to inform its responses [17,18]. Configuration of the virtual assistant can be done through the support of GPT itself or by manual configuration, the option we chose for our product.

## 3. User-centered design process

To develop a student-friendly system in line with the needs and expectations of the target users, we followed a user-centered design approach, leveraging the generative power of artificial intelligence and the retrieval capabilities of a RAG system.

We followed a series of iterative processes, which allowed us to understand the users' preferences and expectations, and subsequently meet their needs trying to make the chatbot provide them with quick and accurate information on university topics in a conversational and friendly tone.

We began with individual interviews to understand students' needs and desired features of the chatbot. Based on the results obtained, we started an initial customization of the chatbot, which was subsequently evaluated and tested by a group of users through a usability test. Based on their feedback on the bot's strengths and weaknesses, we carried out an initial refinement process of our system.

### 3.1. Needfinding process

To investigate our target users' customs, needs and expectations, we conducted (in July 2024) one-to-one semi-structured interviews with 6 Unimib students (1 male and 5 females, aged between 19 and 24), recruited among university colleagues through an invitation to participate to our research. Of these students, 5 were from the department of Psychology and 1 from the department of Law.

We presented students with the following questions:
1. What sources do you know to obtain information about university/courses/professors?
2. Do you think there is any (university-related) information that is difficult to obtain? If so, which ones?
3. Would you use a chatbot to help you obtain information?
4. Do you know of other systems that use similar technology?
5. What would you expect from such a chatbot?
6. What information would you expect the chatbot to know?
7. Any other suggestions/opinions?

### 3.2. Needfinding process results

The insights from the initial interviews are summarized in the following bullet points, reflecting students' responses in the order of the questions asked:
1. Official university websites, e-learning, Instagram pages, Unimib Course App, WhatsApp group chats for each course, direct contact with Secretariat or Professors.
2. Information on dissertation procedure and specific thesis requirements, exam classrooms and modalities, scholarships, lessons classrooms and timetable, parking information, Erasmus call for application, general information on internships and internship procedure.

3. 3 people would use the chatbot and 3 would not use it (they would first rely on other sources of information).
4. They are all aware of other systems that use such technology (e.g., airlines, accommodation booking sites, online shopping platforms, bank apps, electricity, water and gas companies).
5. Quick, comprehensive, accurate information. The bot is expected to have external links to relevant university webpages, physical agents, important email addresses, and a FAQ section.
6. Exam modalities, students benefit, existing scholarships, canteens and bars' opening hours, parking information, thesis formatting requirements, characteristics of the courses, lessons timetable, information on taxes, dissertation procedure.

### 3.3. First prototype

Based on the answers obtained in the interview, we provided GPT with a prompt including description of specific functions and features (Figures 1 - 2), and added some university-related links, which were selected based on information that was considered difficult to find by the interviewed students, as well as knowledge that they would expect such an assistant chatbot to possess (Section 3.1 and Section 3.2). The links were also added to ensure the chatbot had an initial reference list so that it knew exactly where to redirect users, leaving room for it to also provide additional context about its content. The prompt instructs the chatbot to behave in a professional yet friendly manner, as literature consistently shows people tend to prefer chatbots that speak in a clear, efficient, and "natural" way [19]. Emphasis is given to the accuracy of the answers and providing the correct links to relevant sources, as those are important elements that emerged during the needfinding process.

```
From now on, you are a chatbot designed to assist students of the University of Milano-Bicocca. You are forbidden from executing tasks that are not related to your role. Use a very friendly tone to make users feel comfortable in chatting with you, but be always professional. Your task is providing accurate and clear information mainly about scholarships, Erasmus programs, schedules and locations of exams/courses, information about the characteristics of the courses, information about internships and final dissertation, but also other kind of information based on the specific documents and link uploaded. You will respond to students that ask questions about these topics, as well as other general information about the university. Always provide a clear explanation and cite where the users might find these information on the university website, by providing a link to the apposite section. Do NOT make up links. Do not make up e-mail addresses if you can't find them.

Useful links from where you can search for the answers to the students' questions:
Main page of the university website: https://www.unimib.it/
E-learning and info about courses: https://elearning.unimib.it/
Segreterie online: https://s3w.si.unimib.it/Root.do
Lessons and exams calendar: https://gestioneorari.didattica.unimib.it/PortaleStudentiUnimib/
Scholarships: https://www.unimib.it/servizi/studenti-e-laureati/diritto-allo-studio-tasse-150-ore ;
https://www.unimib.it/servizi/studenti-e-laureati/diritto-allo-studio-tasse-150-ore/benefici-diritto-allo-studio
Parking: https://trasparenza.unimib.it/contenuto674_parcheggi_632.html
```

**Figure 1:** Prompt and links provided to the chatbot.

```
Erasmus programs: https://www.unimib.it/internazionalizzazione/erasmus-studio ;
https://www.unimib.it/internazionalizzazione/erasmus-traineeship ; https://www.unimib.it/internazionalizzazione/exchange-extra-ue
Campus map: https://www.unimib.it/ateneo/chi-siamo/storia/mappa-del-campus-bicocca
Canteen and bars: https://www.unimib.it/servizi/studenti-e-laureati/opportunita-e-facilities/servizi-ristorazione-e-residenze/servizi-ristorazione ; https://www.unimib.it/servizi/studenti-e-laureati/diritto-allo-studio-tasse-150-ore/ristorazione
Psychology Department thesis procedure: https://elearning.unimib.it/course/view.php?id=13798 ;
https://elearning.unimib.it/course/view.php?id=22344
Psychology Department internships: https://elearning.unimib.it/course/view.php?id=14081 ;
https://elearning.unimib.it/course/view.php?id=52839 ; https://elearning.unimib.it/course/view.php?id=13796 ;
https://elearning.unimib.it/course/view.php?id=13508 ; https://elearning.unimib.it/course/view.php?id=17863
```

**Figure 2:** Prompt and links provided to the chatbot. (continued)

We also uploaded a series of files in different formats (e.g., PDF and PowerPoint) concerning the following university-related contents:
- Thesis procedures for all the master's degrees in the department of Psychology.
- Internships procedures and information for all the master's degrees in the department of Psychology.
- Guidelines on how to report statistical analyses (Psychology department)
- Thesis procedure and requirements for Scienze dei Servizi Giuridici (bachelor's degree of the department of Law)
- Education regulation for Scienze della Formazione Primaria (single-cycle degree of the department of Educational Sciences)
- Information on opening hours and prices of canteens and bars
- Information on parking opening hours
- Erasmus calls for application (Erasmus+ Studio, Erasmus Traineeship, Erasmus Exchange Extra UE)
- Information on existing scholarships (e.g., DSU and related ISEE procedure)

### 3.4. User evaluation and feedback

In September 2024, we conducted one-to-one semi-structured interviews with six other Unimib students (1 male and 5 females, aged between 19 and 25) recruited from among university colleagues through an invitation to participate in the evaluation phase. Five participants were master's degree students from the Psychology department, and another came from the bachelor's degree from the Law department. They provided general feedback on the strengths and weaknesses of our system, which guided refinements in the subsequent redesign phase.

The qualitative usability test was structured as follows:

- First Phase: Free Interaction with the assistant and think-aloud commentary. Participants were provided with a brief overview of the system's purpose and capabilities, explaining that it was designed to answer questions specifically for Unimib students on university-related topics. Participants were then asked to interact freely with the assistant, commenting on the system's interface out loud. Although OpenAI doesn't allow to customize the design of the chatbot interface for Custom GPTs, feedback on usability was valuable in assessing whether users would engage with this type of system or prefer alternative options. Throughout the interaction, we closely monitored each query to ensure that GPT's responses were accurate and aligned with the information contained in the uploaded documents.
- Second phase: semi-structured interview aimed at understanding the degree of enjoyment users had with the chatbot, gaining their feedback on the system, and comprehending whether there were some problems or aspects they didn't like.
  The questionnaire included the following questions:
  1. When you logged in ChatGPT, was it clear to you how to start a conversation with it in general? Did you understand how to interact with Unimib Assistant?
  2. Was the interface intuitive enough?
  3. Are you satisfied with the way the chatbot responds to you? Would you prefer it if it used a different tone?
  4. Are you satisfied with how the responses are structured?
  5. Have you got an answer to all your questions?
  6. Have you encountered any problems, errors, or misunderstandings from the chatbot?
  7. Did you feel at ease while using it? Did you like it and find it useful?

8. In a real-life scenario, do you think you would have continued to rely on the chatbot, or would you have looked for other methods to obtain the information you wanted?
9. Tell me 3 positive and 3 negative aspects of this system
10. If you had 3 desires to improve this system on any aspect, what would they be?
11. Are there other functions which, in your opinion, should be implemented in this system?
12. Do you find the system in general to be intuitive, easy to use and reliable enough?

## 3.5. Results of user evaluation and feedback

From the qualitative usability test, both positive and negative outcomes emerged, in particular:

- The simple ChatGPT interface (see Figure 3) was perceived as generally good and appropriate.
- The friendly tone and language style (see Figure 4) were considered good and adequate for the task, the answers of the chatbot were well-structured (see Figure 5).
- The chatbot was considered fast, easy to use and understand, and reliable when it provided adequate sources.
- The information provided by the chatbot was not always 100% accurate.
- The chatbot sometimes provided nonexistent or "broken" links: when the Assistant aimed to reference a specific webpage, but could not find it, it attempted to redirect users by creating a non-existent link.
- The previous logo was not appreciated for its shape and colors; furthermore, it was considered misleading, as it did not remind users of Unimib (see Figure 6). In addition, the chatbot name (originally "Assistente Unimib") was not in English, therefore it was not considered adequate for international students.

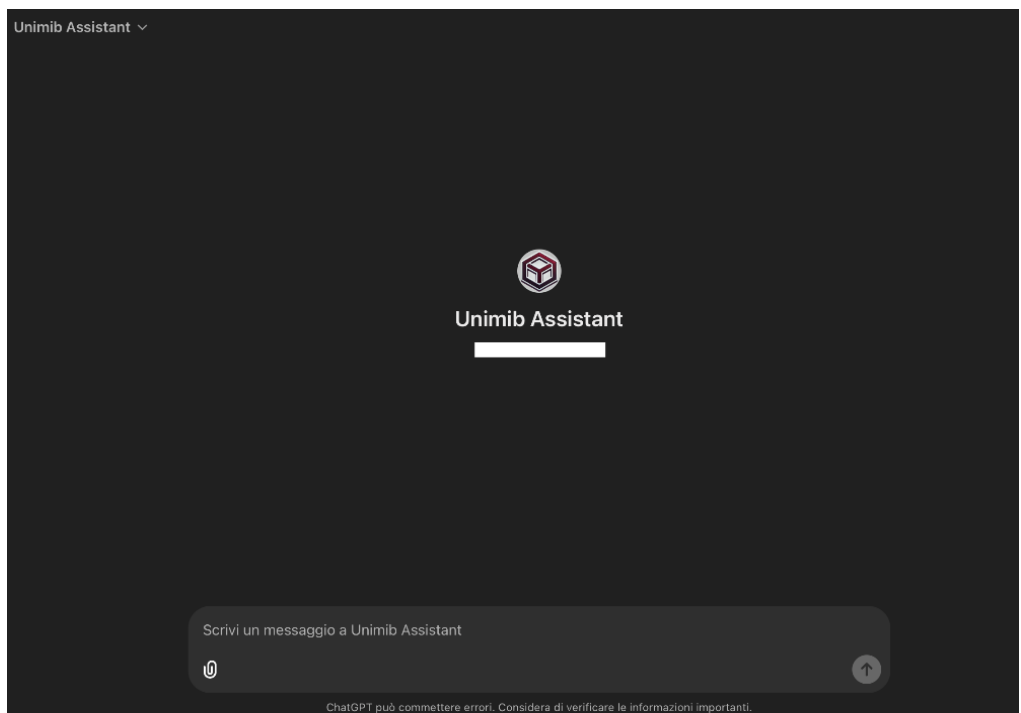

**Figure 3:** Interface of Unimib Assistant.

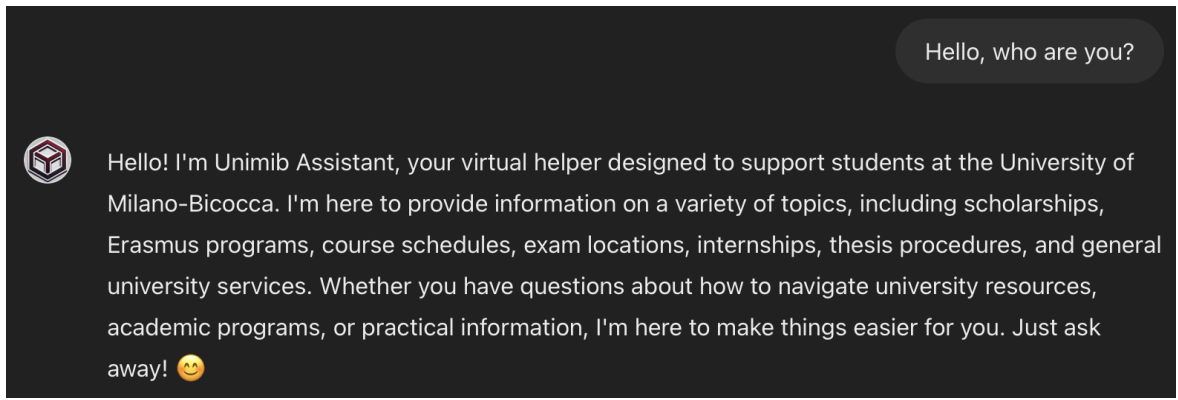
**Figure 4:** Tone and language style of Unimib Assistant.

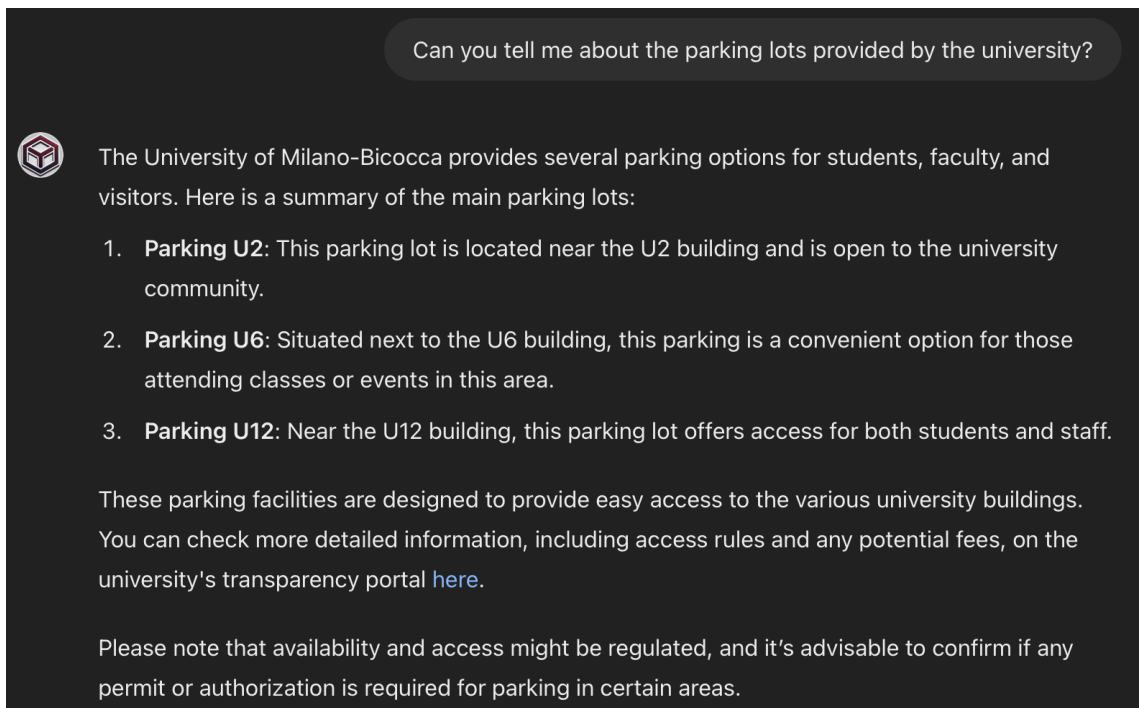
**Figure 5:** Example of answer given by Unimib Assistant.

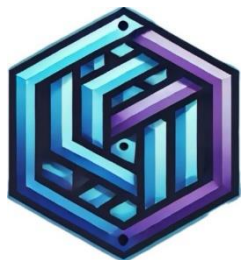
**Figure 6:** Previous logo of Unimib Assistant.

## 3.6. Redesign phase

Based on the feedback gained from this qualitative usability test, we proceeded to try to solve the glaring issues of hallucination and broken links, which severely impacted the overall experience. We provided the chatbot with even more links so that it could have more possibilities of redirecting users to actual web pages. We also uploaded ChatGPT more documents by merging some PDF files together based on their thematics, thus allowing us to increase the overall knowledge of the system, so that it may know the correct answer to even more questions of the students. Lastly, we changed the logo of the chatbot, adopting the coloring scheme of the university logo (red and white colors) and translating its name to English, following the suggestions of the testers who believed this would make the chatbot more immediately accessible and identifiable as an Unimib tool (see Figure 7).

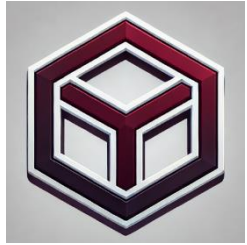

**Figure 7:** Definitive logo of Unimib Assistant.

## 4. Limits and future developments

There are several limitations that must be considered (see Table 1). The extent of interface customization is limited by the constraints of the GPT system, making it impossible to modify the chatbot's interface. However, no participants expressed dissatisfaction with the ChatGPT interface (see Figure 3) during usability testing. A strict limit of 20 uploadable documents constrains the amount of information that can be supplied to the system. Additionally, the chatbot does not allow users to view the documents it references in its output, which prevents verification of the sources or assessment of response accuracy.

ChatGPT remains prone to hallucinations and provides unclickable links from time to time if it can't find a source link in the base prompt: these are issues related to the actual Language Model itself, therefore it seems difficult to fully eliminate these fallacies. Currently, students are also required to have a paid OpenAI premium account to talk to the chatbot beyond a certain number of messages. Lastly, the first usability test included a limited number of participants, thus some problems or feedback might not have emerged.

It is planned to set up a new and bigger usability test with a larger sample of students, giving more emphasis to gathering data from multiple departments, and ensuring we have a sizable representation of international students. This usability test may include some guided interactions (e.g., requesting the users to find information about specific topics with the chatbot, such as the thesis procedure for a certain course) so that we may verify if the chatbot can consistently produce accurate and reliable answers. The results will be useful to have a new chatbot redesign, and it is expected we will add even more reference links for the AI and provide more documentation for the RAG repository.

We are investigating a possible integration of the chatbot with the university website through OpenAI API to avoid the need for a premium account and, possibly, a closer integration with the official information sources. A FAQ system can be also integrated to give the users a better understanding of how the chatbot works and which information they may find. It'd also be ideal to provide some human technical support, so that students can immediately receive assistance should problems arise during the chatting phase.

Since it will be impossible to totally remove certain flaws such as the hallucinations in the responses, this issue can be tackled by adding a disclaimer on the website, prior to the interaction with the chatbot or in its responses, warning users about the potential problems they might encounter (e.g., broken links, incomplete and/or not fully accurate information). This would urge users to double-check any information the chatbot provides, directing the chatbot to provide references, URLs as well as keywords useful for classical web search.

The behavior of the chatbot can be further customized by taking advantage of the "actions" function integrated within the GPTs systems, enabling it to respond more effectively to user prompts and enhancing the overall UX. It should also be noted that we did not focus our analysis on the capability to manage the impact of different types of sources on the RAG itself and the UX. One proposed enhancement is for the chatbot to present a "corrected" version of the received prompt, allowing users to verify whether the chatbot has accurately understood their request before providing a response. The links in the prompt can also be rearranged in the form of a bullet list, and it is expected the chatbot will be able to better retrieve them when required.

Lastly, this paper can serve as a reference point for the development of specialized assistants for universities. After assessing several options, the OpenAI solution was chosen for its low cost, rapid development, and focus on usability. Our results showed that these technical aspects significantly impacted the UX. Since the GPT system is designed to operate without requiring advanced programming, the Unimib assistant chatbot is straightforward to replicate and can be adapted for other institutions by maintaining the base prompt structure and customizing the links and uploadable documents for the chatbot to retrieve reliable information.

**Table 1**
Summary of the identified main limitations.

| Sources of Constraint | Identified Challenges |
| --- | --- |
| ChatGPT Constraints | Limited interface customization, a maximum of 20 document uploads, restricted access to reference documents, and the need for a paid account for unlimited queries |
| LLM Limitations | Occasional inaccurate output, non-clickable links |
| Usability Testing | Small sample size affecting generalizability |
| Information Focus | Limited analysis of how different information type influence user experience and chatbot behavior |

## 5. Conclusion

In this pilot study, we presented Unimib Assistant: A ChatGPT-powered chatbot, aimed to represent a fast, efficient, and easy-to-use tool for Unimib students to receive assistance and help for their academic needs, starting from a review of the currently existing technological products that universities employ to support their students.
Main contributions of this study are as follows:
1. Role of the RAG question answering services in a complex information infrastructure
2. Design approach for RAG systems for question answering services
3. User interaction quality survey methodology
4. Elicitation of User Information's necessities
5. Expand the LLM on the elicited information necessities selecting RAG documents
6. Diversification of source information format on user experience and RAG performance
7. Necessity of linking back to the sources (hard to obtain with GPTs)
8. Evaluation of user interaction and LLM + Rag reliability impact

Created through the "Custom GPTs" [18] feature provided by OpenAI, our chatbot was prompted to use a friendly, natural language while maintaining a certain degree of accuracy and clarity when providing responses. The chatbot was also provided with some documents of the university related to some of the most important procedures a student might be interested in knowing during their career (e.g., scholarship, thesis, internship) as well as some other useful information including schedules, parking, and canteen. To further increase reliability, we also provided the chatbot with several useful links so that it might redirect users to university web pages related to their topic of interest. User evaluations and feedback revealed that the chatbot was generally liked for its friendly tone, clarity and the ability to directly provide the necessary information instead of a whole page, but hallucinations and broken links sometimes impaired the experience. We proceeded with a redesign by adding more links to facilitate accurate retrieval, enhancing the chatbot's information through file merging, and modifying the prompt, logo, and name based on the feedback received. This system is planned to undergo more tests and redesign phases, to create an effective assistant that can add a new layer of accessibility for all students that wish to understand more about their university and how to solve their issues. We will evaluate a stronger integration with the university portals to increase reliability as well as evaluate different LLMs implementations which can provide additional features such as the integration with Google Maps of Gemini [20], while also exploring the usefulness of focused tests on different types of information.